\documentclass[12pt]{article} 

\input epsf

\begin{document}

\begin{center} {\Large \bf Strategy for investments from Zipf law(s) } 

\vspace*{1.5cm}

{\bf M. Ausloos\footnote{e-mail address: marcel.ausloos@ulg.ac.be} and Ph. Bronlet}

\vspace*{1.5cm}

SUPRAS and GRASP, B5, Sart Tilman Campus, B-4000 Li$\grave e$ge, Euroland 

\date{\today}


\vspace*{1.5cm}


\end{center} \begin{abstract} { We have applied the Zipf method to extract the $\zeta'$ exponent for seven financial indices (DAX, FTSE; DJIA, NASDAQ, S\&P500;
Hang-Seng and Nikkei 225), after having translated the signals into a text based
on two letters. We follow considerations based on the signal Hurst exponent and the notion of a time dependent Zipf law and exponent in order to implement two simple investment strategies for such indices. We show the time dependence of the
returns.}

\end{abstract} \vskip 0.5cm


\vskip 0.5cm

\section{Introduction}

Uusally analysts recommend investment strategies based e.g. on "moving averages",
"momentum indicators", and the like techniques.\cite{Fama,Apfa3} As soon as econophysicists discovered scaling laws in financial data, it was of interest to
search for some predictive value from the laws through some extrapolated evolution. E.g. a technique known as {\it detrended fluctuation analysis} (DFA) which measures the deviation of correlated fluctuations from a trend was developed into a strategy known as the $local$ (or better $instantaneous$) DFA in
order to predict fluctuations in the exchange rates of various currencies, Gold price and other financial indices. \cite{1,2} The statistical analysis of data was based on the value of the exponent of the so found scaling law, itself related to the fractal dimension of the signal, or also to the Hurst exponent of
the so called rescaled range analysis. Mathematical extensions, so called $q-order$ DFA and multifractals, can be found in the literature, though optimization problems and predictions on the future of fluctuations are apparently not so evident from these methods. A drawback in the DFA is found in the fact that it rather looks at correlations in the sign of fluctuations rather
than at correlations in their amplitude. 

Another sort of data analysis technique is known as the Zipf technique,\cite{Zipf,West} originating in work exploring the statistical nature of languages. The Zipf analysis technique has also been used outside linguistic,
financial and economic fields.\cite{Zipfoutside} The technique is based on a {\it
Zipf plot} which expresses the relationship between the frequency of words (more
generally, events) and the rank order of such words (or events) on a log-log diagram; a $cumulative$ histogram can be drawn as well. The slope of the best linear fit on such a plot corresponds to an exponent $s$ describing the frequency
$P$ of the ($cumulative$) occurrence of the words (or events) according to their
rank $R$ through, e.g. $P(>R) \simeq R^{-s}$. 

There are many instances in which financial and other economic data can be described through a log-log (Zipf) plot : e.g., the distribution of income (Pareto distribution) \cite{Pareto}, the size of companies \cite{companysize}, sociology \cite{MarsiliZhang}, sometimes after translating the financial data into a text
\cite{VandewalleAusloos,Okuyama,Takayasu,IvanovaAusloos2,AusloosLadek}. Thus it seems of interest to check whether such a technique can have some predictive value in finance. The present report is in line with such previous investigations. We present results based on considerations that financial data series are similar to fractional Brownian motion-like time series, and usually biased.\cite{IJMPC} We examine whether a time dependent Zipf law and exponent exist and can be used in order to implement simple investment strategies. 

First, it is thus necessary to translate the financial data into a $text$, based
on an alphabet with $k$ characters and search for word s of length $m$. There are
obviously $k^m$ possible words. They can be ranked according to their frequency on a log(frequency)-log(rank) diagram. A linear fit leads to consider the relationship as a power law. Moreover, in the spirit of the local DFA, a local (or ''time'' dependent) Zipf law or exponent can be introduced.\cite{IJMPC} In this latter reference, we have also considered the effect of a linear trend on the value of the Zipf exponent.

\begin{table}[tbh] \begin{center} \caption{ Typical financial indices characteristics between Jan. 01, 1997 and Dec. 31, 2001: Hurst exponent $H$, (5,2)-Zipf exponent, bias ($\varepsilon$), $p_u$, $p_d$, linear trend slope} \tabcolsep=4pt
\begin{tabular}{lccccccc} \hline & $H$ & $\zeta_{(5,2)}$ & $\varepsilon$ & $p_u$
& $p_d$ & Trend \\ \hline 1. DAX	& 0.51$\pm$0.01 & 0.11$\pm$0.05 & 0.0332 &
0.5332 & 0.4668 & 2.24$\pm$0.07\\ 2. DJIA & 0.46$\pm$0.04 & 0.11$\pm$0.02 & 0.0172 & 0.5172 & 0.4828 & 2.97$\pm$0.07\\ 3. FTSE	& 0.43$\pm$0.03 & 0.19$\pm$0.08 & 0.0149 & 0.5149 & 0.4851 & 0.90$\pm$0.05\\ 4. Hang-Seng & 0.47$\pm$0.02 & 0.08$\pm$0.02 & 0.0060 & 0.5060 & 0.4940 & 1.75$\pm$0.20\\ 5. Nasdaq & 0.56$\pm$0.03 & 0.19$\pm$0.08 & 0.0428 & 0.5428 & 0.4572 & 1.19$\pm$0.06\\ 6. Nikkei225 & 0.47$\pm$0.06 & 0.15$\pm$0.04 & -0.0204 & 0.4796 &
0.5204 & -4.69$\pm$0.16\\ 7. S\&P	& 0.51$\pm$0.04 & 0.12$\pm$0.03 & 0.0152 &
0.5152 & 0.4848 & 0.38$\pm$0.01\\ \hline \end{tabular} \label{tab 2.1} \end{center} \end{table}

Here below we have translated seven financial index signals (DAX, FTSE; DJIA, NASDAQ, S\&P500; Hang-Seng and Nikkei 225) each into a text based on two letters
$u$ and $d$. Based on the above considerations we have imagined two simple investment strategies, and report on the results (or "returns"). From the beginning we stress that a restriction to two letters is equivalent to examine only correlations in the fluctuation signs. However the Zipf method main interest
is surely the capability to consider amplitude fluctuations, - by defining various fluctuation ranges.

\section {Data analysis}

The daily closing values of (DAX, FTSE; DJIA, NASDAQ, S\&P500; Hang-Seng and Nikkei 225) indices, from Jan. 01, 1997 till Dec. 31, 2001 (Fig.1) have been obtained from $ http://finance.yahoo.com/ $. They contain $ca. $1250 data points.
After translating the financial time series into a text, one searches for words,
and rank them according to their frequency. On a log-log paper, the best line fit
slope is the Zipf exponent. Elsewhere we have already shown that the usual Zipf exponent $\zeta $ \cite{Zipf,West} depends on the normalization process used to calculate the ranks. If the frequency $f$ of occurrence is normalized with respect to the theoretical one $f'$ , i.e. that expected for pure (stochastic) Brownian processes, one has $f/f'$ $\sim R^{-\zeta '}$. The theoretical frequency
expected for a letter in a text based on a binary alphabet, $u,d$ takes into account the number $n$ of characters, say of type $d$ ( and $u$), in a word. Suppose that in the text, the frequency of a $d$ ($u$) letter is $p_d$ ($p_u$). Usually, a bias exists, i.e. $p_{u} \neq p_{d} $. Therefore $f'=p_{u}^{(m-n)}p_{d}^{(n)}$. Whether or not the $\zeta$ and $\zeta'$ exponent depend on the bias has been examined elsewhere.\cite{IJMPC} The $p_u$ and $p_d$ values for the seven indices are reported in Table 1, together with the bias defined here as $\epsilon = p_u - 0.5$. The linear tendency for the time interval
is also given in Table 1. We have calculated overall Zipf exponent values, $\zeta_{(m,k)}$, and give the $\zeta_{(5,2)}$ value for the seven indices in Table 1.

In the spirit of the so called local (or better $instantaneous$) DFA method. We can consider that a Zipf exponent is time dependent, thus obtain a $local$ Zipf law and $local$ Zipf exponent. Only the case for $m <$8 letter words has been considered, but are not shown for lack of space. This $m$ value is so chosen within the financial idea background having motivated this study, e.g. $m=5$ is the number of days in a (bank) week !

In general a (one dimensional) financial index can be characterized by a so called Hurst exponent $H$, obtained as follows. The time series is divided into boxes of equal size, each containing a variable number of ''elements''. The local
fluctuation at a point in one box is calculated as the deviation from the mean in
that box. The $cumulative$ departure up to the $j^{th}$-point in the box is next
calculated in all boxes. The rescaled range function is next calculated from the
difference between the $maximum$ and the $minimum$, i.e. the range in units of the rms deviation in the box. The average of the rescaled range in all boxes with
an equal size $n$ is next obtained and denoted by $<R/S>$. The above computation
is then repeated for different box sizes $s$ to provide a relationship between $<R/S>$ and $s$, - which is expected to be a power law $<R/S>\simeq s^{H }$ if some scaling range and law exist.

If $H=1/2$ one has the usual Brownian motion. The signal is said to be persistent for $H>1/2$, and antipersistent otherwise. We have calculated the H urst exponent
\cite{Hurst} by this rescaled range analysis \cite{Addison} for the seven financial index signals. Their $H$ value and the corresponding error bar are given in Table 1. The error bars are those resulting from a best linear fit and a
root mean square analysis.

\begin{figure} \begin{center} \leavevmode \epsfysize=4cm \epsffile{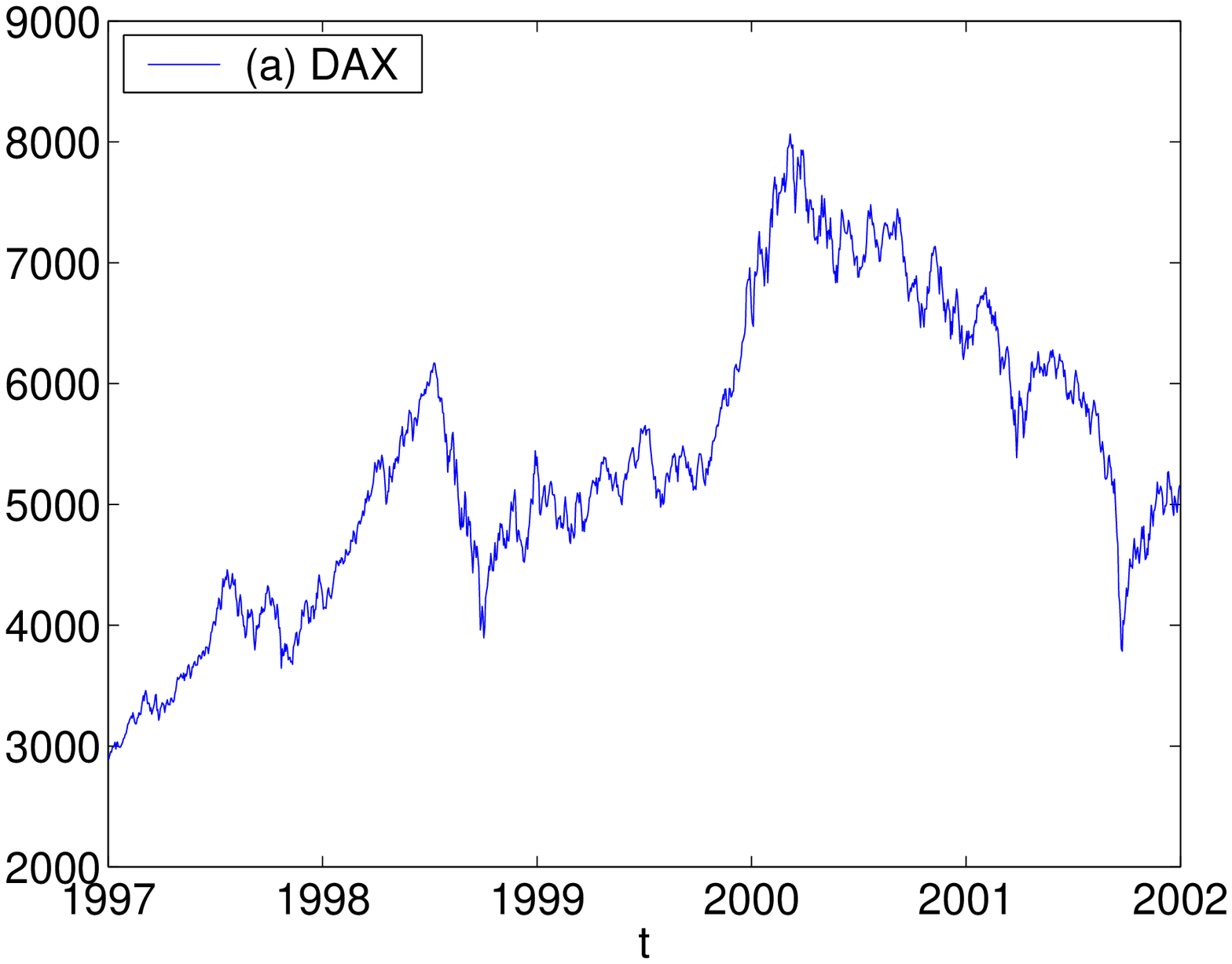} \hfill \leavevmode \epsfysize=4cm \epsffile{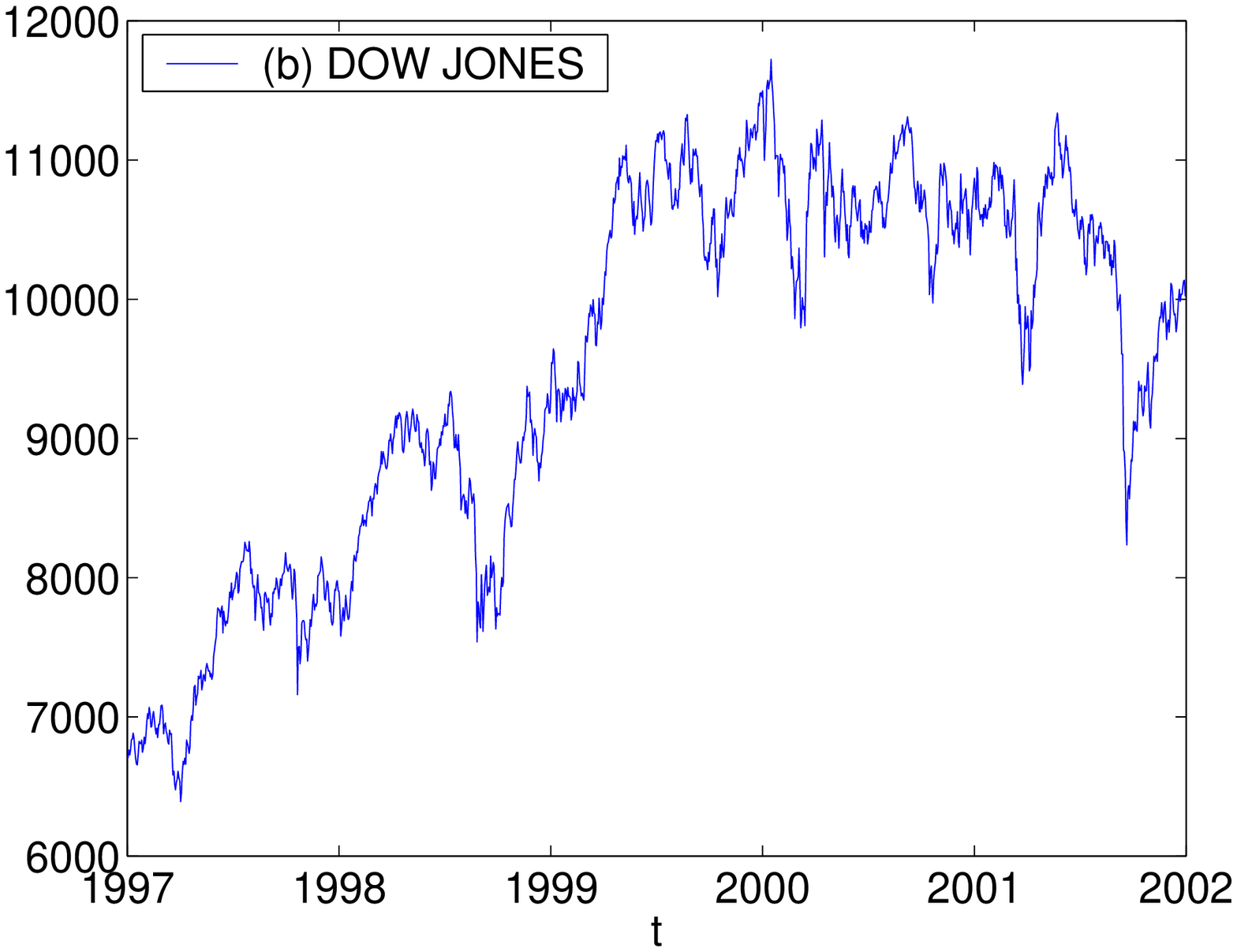} \vfill \leavevmode \epsfysize=4cm \epsffile{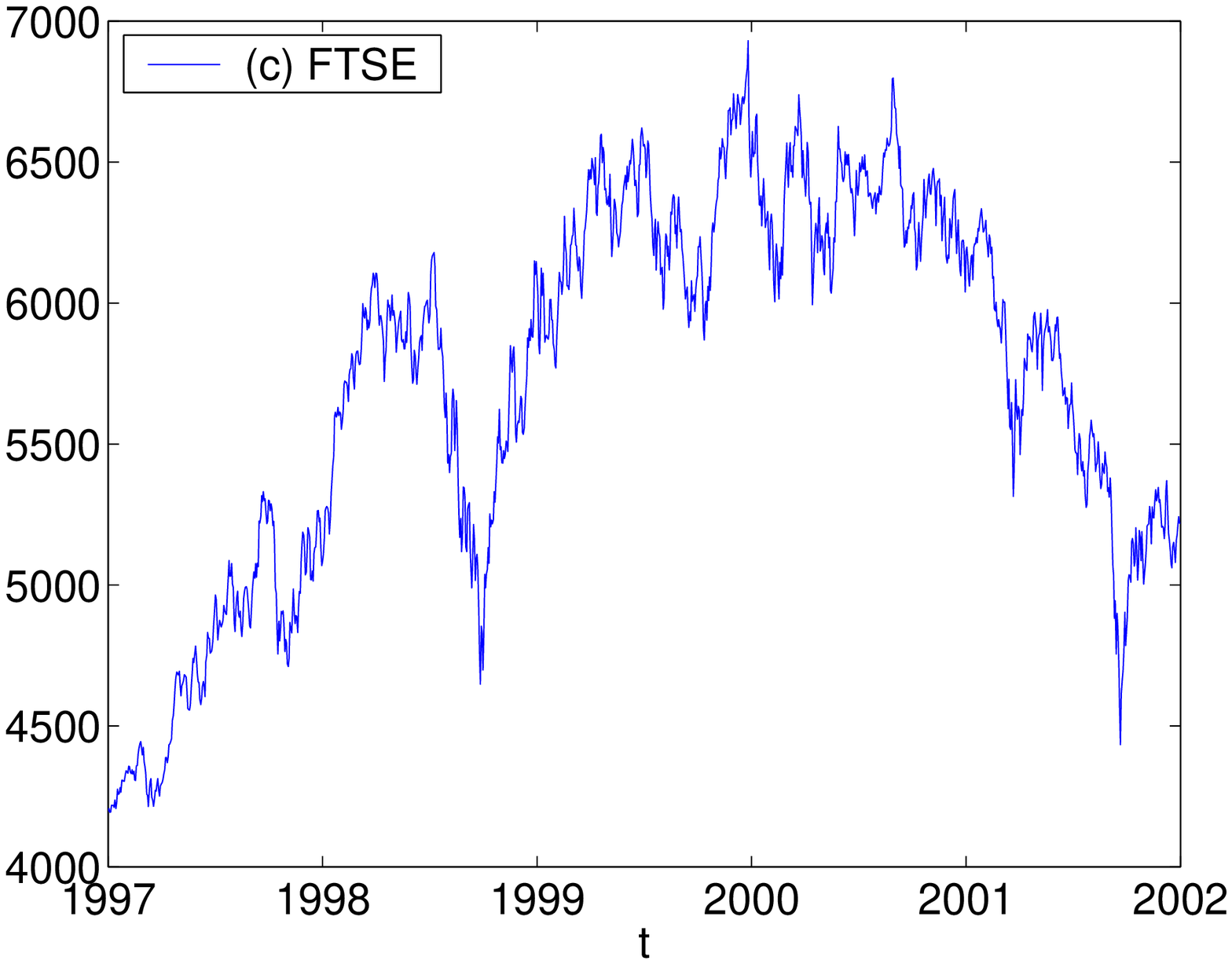} \hfill \leavevmode \epsfysize=4cm \epsffile{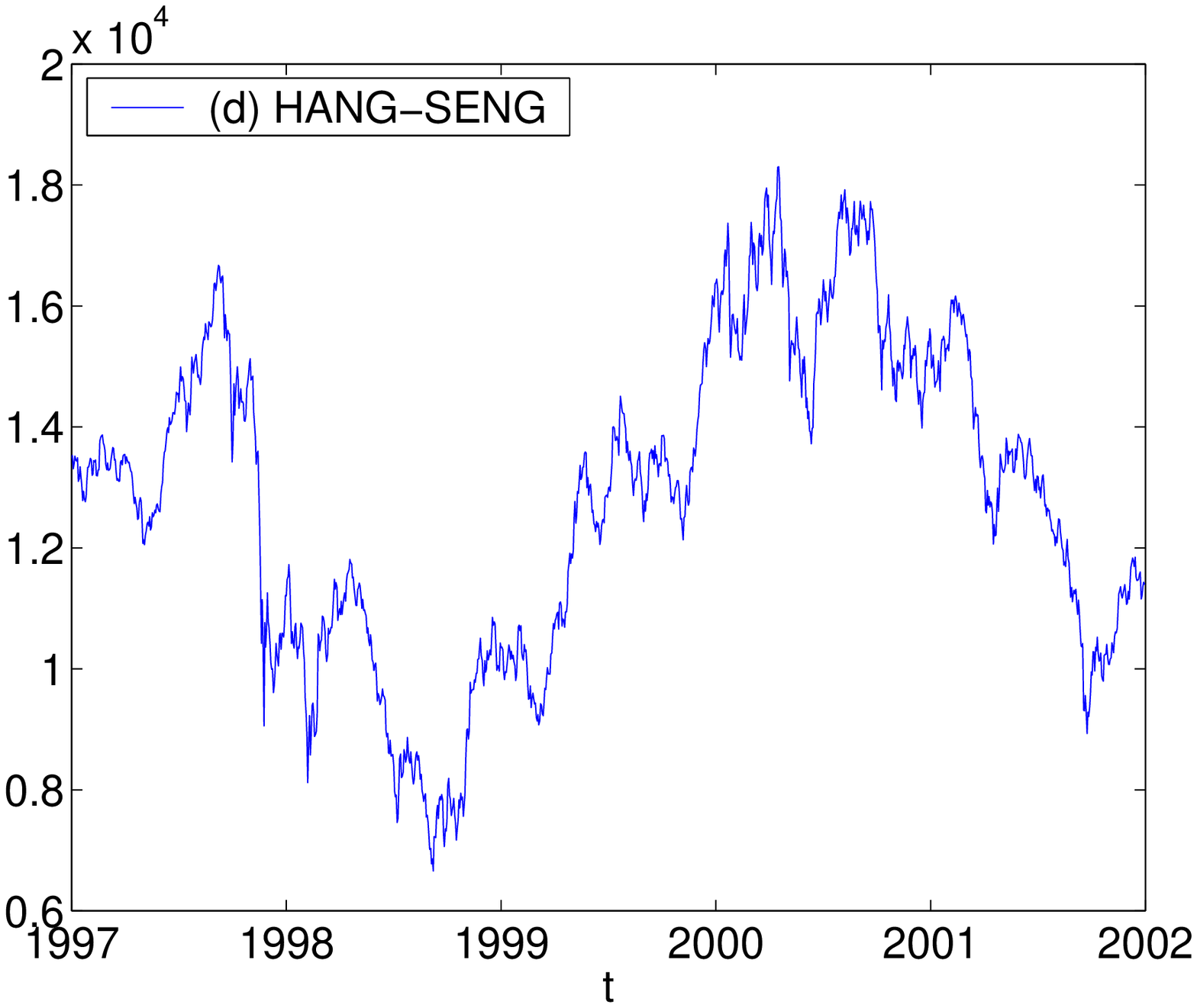} \vfill \leavevmode \epsfysize=4cm \epsffile{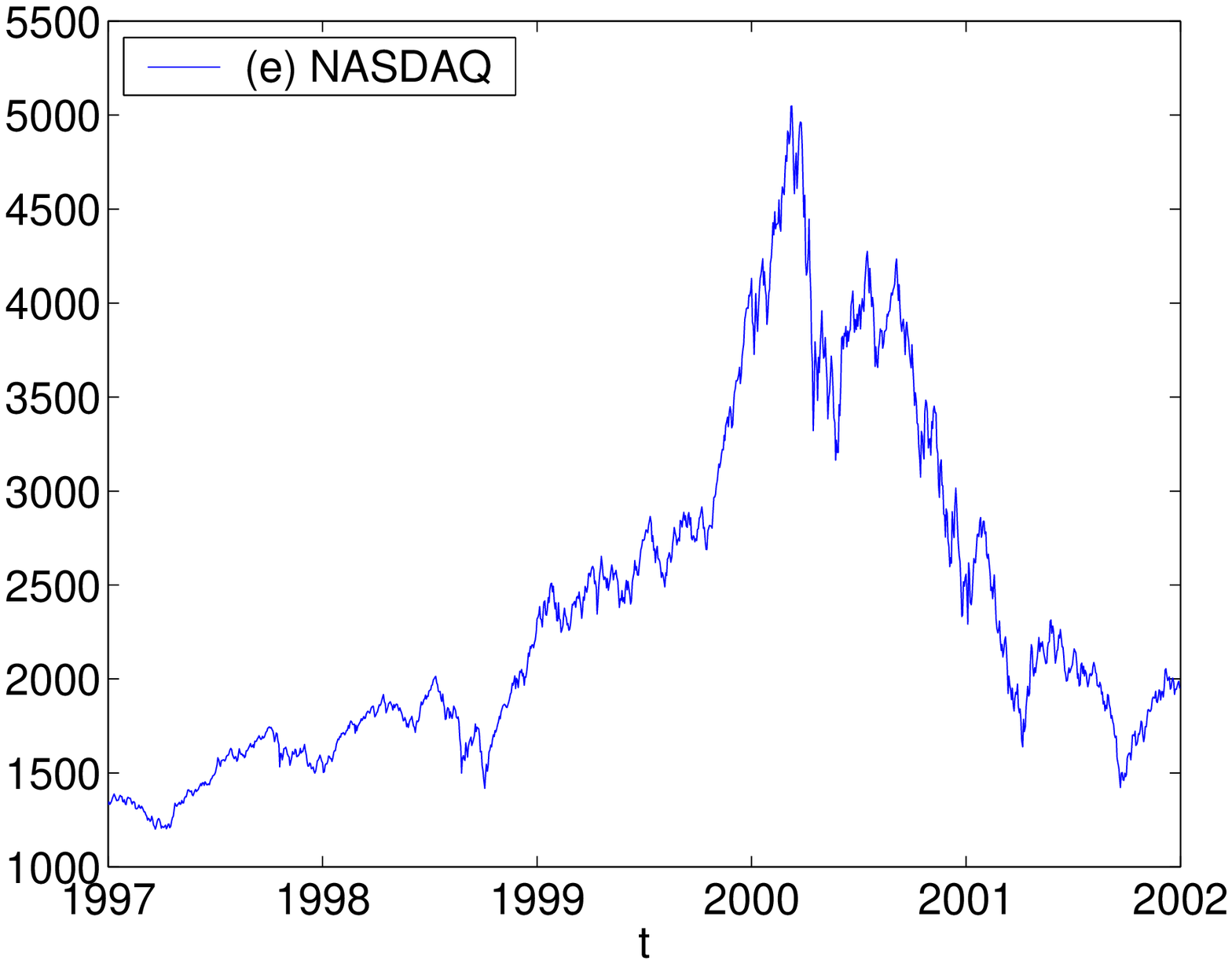} \hfill
\leavevmode \epsfysize=4cm \epsffile{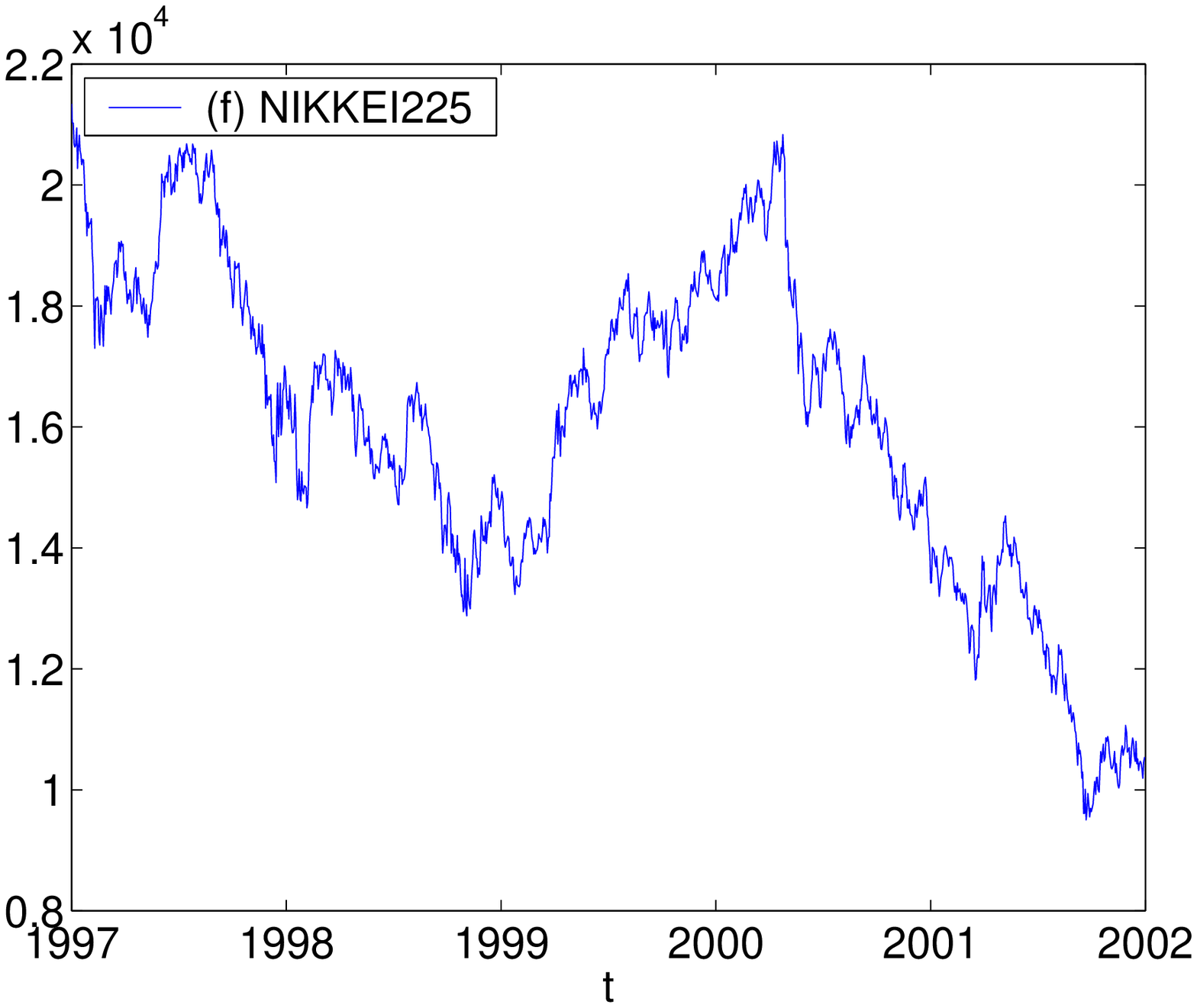} \vfill \leavevmode \epsfysize=4cm
\epsffile{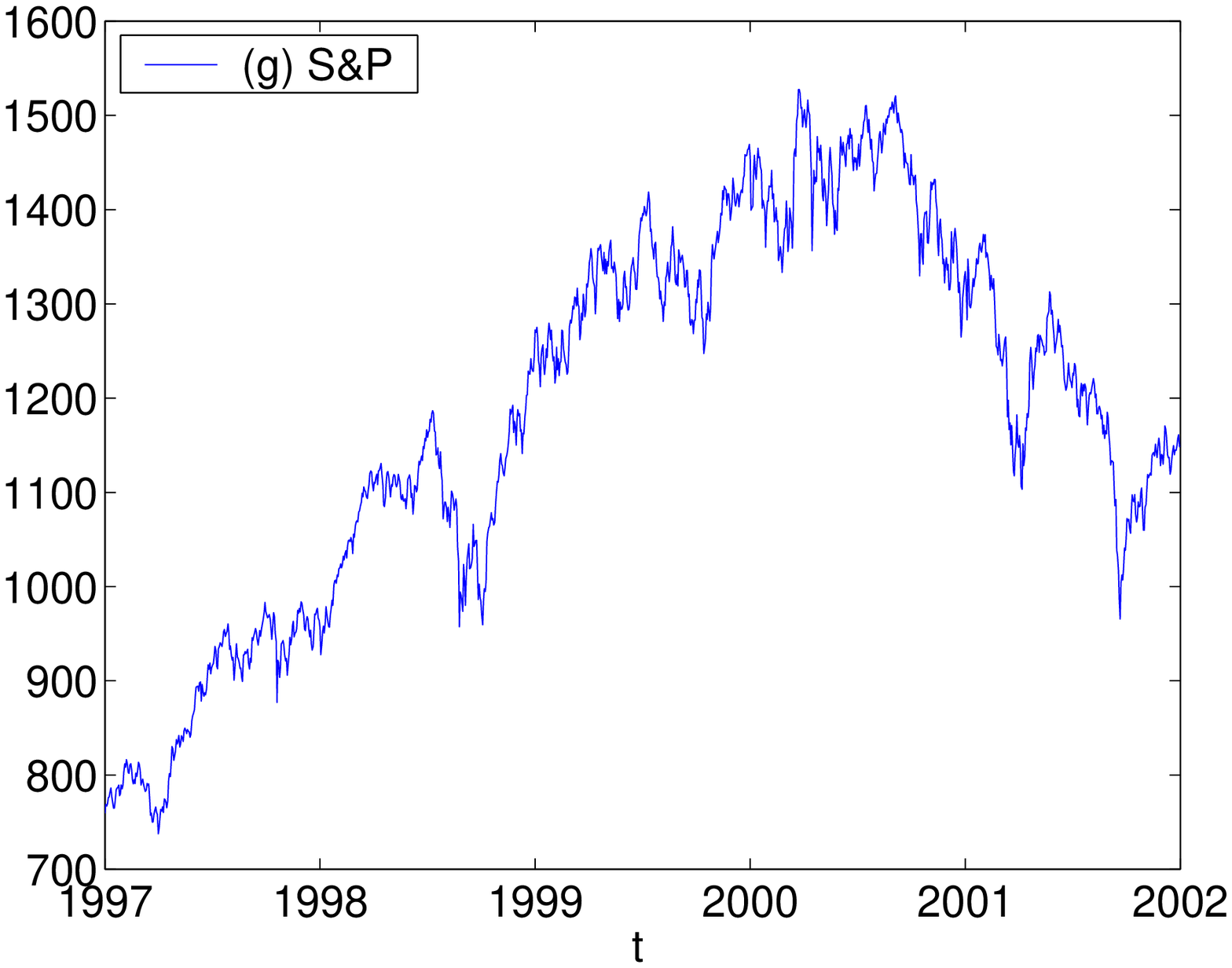} \caption{The (DAX, FTSE; DJIA, NASDAQ, S\&P500; Hang-Seng and
Nikkei 225) indices have been obtained from $http://finance.yahoo.com/$. They contain $ca. $1250 data points, from Jan. 01, 1997 till Dec. 31, 2001} \end{center} \end{figure}

Tests (not shown here) of the stochasticity (or not) of the data can be based on
the surrogate data method \cite{surrogate} in which one randomizes either the sign of the fluctuations or shuffles their amplitude, and finally observes whether the error bars (or confidence intervals) of the raw signal and the surrogate data signal overlap.

\section{Returns and basic Zipf strategy} 

The method is based on searching for the probability of a character sequence at the end of a word. We consider the case of what can happen the {\it next day} after a few ($m-1$) days) only. Consider a word of length $m-1$, and calculate in
all boxes of size $\tau$ the probabilities $p_u(t)$ and $p_d(t)$ to have a character sequence ($c_{t-m-3}, ...,c_{t-1}, u$) and ($c_{t-m-3}, ...,c_{t-1},d$)
respectively, where $c_t$ represents the character at time $t$. Since only a $k$=2 alphabet is used, it is fair to develop a simple strategy based $only$ on the sign of the fluctuations, thus use a strategy similar to that one implemented
in the "instantaneous" DFA, i.e. when expecting correlated or anticorrelated fluctuations, in $u$ and $d$. In order to avoid investment activity when the choice probability is low we have used a $strength$ parameter for measuring the relative probabilities, i.e.

\begin{equation} K(t) \quad = \quad \mid \frac{ p_u(t) - p_d(t) }{p_u(t) + p_d(t)} \mid , \label{ers} \end{equation} varying between 0 and 1, its value giving the number of shares bought (or sold) at a certain investment time. 

\begin{table}[tbh] \begin{center} \caption{ Final returns $r(t)$ in
(\%) obtained after 5 years on various indices from various strategies $Z1$ and $Z2$ as described in the text when based on a ($m,k$=2) Zipf exponent as compared
(second column) to the mere final index value change} \tabcolsep=5pt
\begin{tabular}{lcccccccc}
\hline $\tau$ = 500	& $r(t)$ & \multicolumn{3}{c}{Zipf1} && \multicolumn{3}{c}{Zipf2} \\ \cline{3-5} \cline{7-9} & & $m=3$ & $m=5$ & $m=7$ && $m=3$ & $m=5$ & $m=7$ \\ \hline 1. DAX & 77.55 & 61.33 & 71.34 & 62.39 && 65.01 & 65.42 & 32.78 \\ 2. DJIA & 49.49 & 47.42 & 55.73 & 29.12 && 34.52 & 39.12
& 23.68 \\ 3. FTSE	& 24.30 & 33.34 & 30.63 & 42.98 && 18.99 & 29.33 & 40.92 \\
4. Hang-Seng & -15.48 & -23.88 & -26.61 & -25.13 && 3.78 & -9.81 & -9.30 \\ 5. Nasdaq & 46.55 & 56.30 & 39.87 & 61.41 && 56.52 & 65.07 & 43.85 \\ 6. Nikkei225 &
-50.61 & -13.78 & -12.39 & -8.73 && -5.47 & -14.60 & -7.89 \\ 7. S\&P	& 51.16 & 67.80 & 80.34 & 107.79 && 43.65 & 57.85 & 71.33 \\ \hline \end{tabular} \label{tab 2.2} \end{center} \end{table} 

Results are reported when windows (boxes) of size $\tau=$500 respectively are moved along the signal. This value corresponds to a 2 year type investment window. Notice that the $local$ exponents are usually larger than the average one, due to finite size effects.

In the $Zipf_1$ ($Z1$) strategy, we consider that if $p_u(t)> p_d(t) $, a "buy order" is given. A ''sell order'' is given for $p_u(t)< p_d(t) $. No order is given when both probabilities are equal. Results reported in Table 2 pertain to $m$= 3, 5, and 7 at the end of the 5 year interval. In the $Zipf_2$ ($Z2$) strategy the local linear trend is subtracted before calculating $p_u(t)$ and $p_d(t)$. The time dependent returns for $Z1$ and $Z2$ in the case $m$=3, 5, and
7, and for $k$=2 are given in Fig. 2 for the seven hereby considered financial indices. A return $r(t)$ (given in $\%$) is defined from\begin{equation} Bq(t) \quad = \quad Bq(t_0) \quad[ 1 + r(t)], \label{ret} \end{equation} where $Bq(t)$
and $Bq(t_0)$ are the amount of money available at time $t$ and at the beginning
$t_0$ of the investment period respectively, for a share of value $q(t)$ bought $q(t_0)$ at the starting date.

\begin{figure} \begin{center} \leavevmode \epsfysize=4cm \epsffile{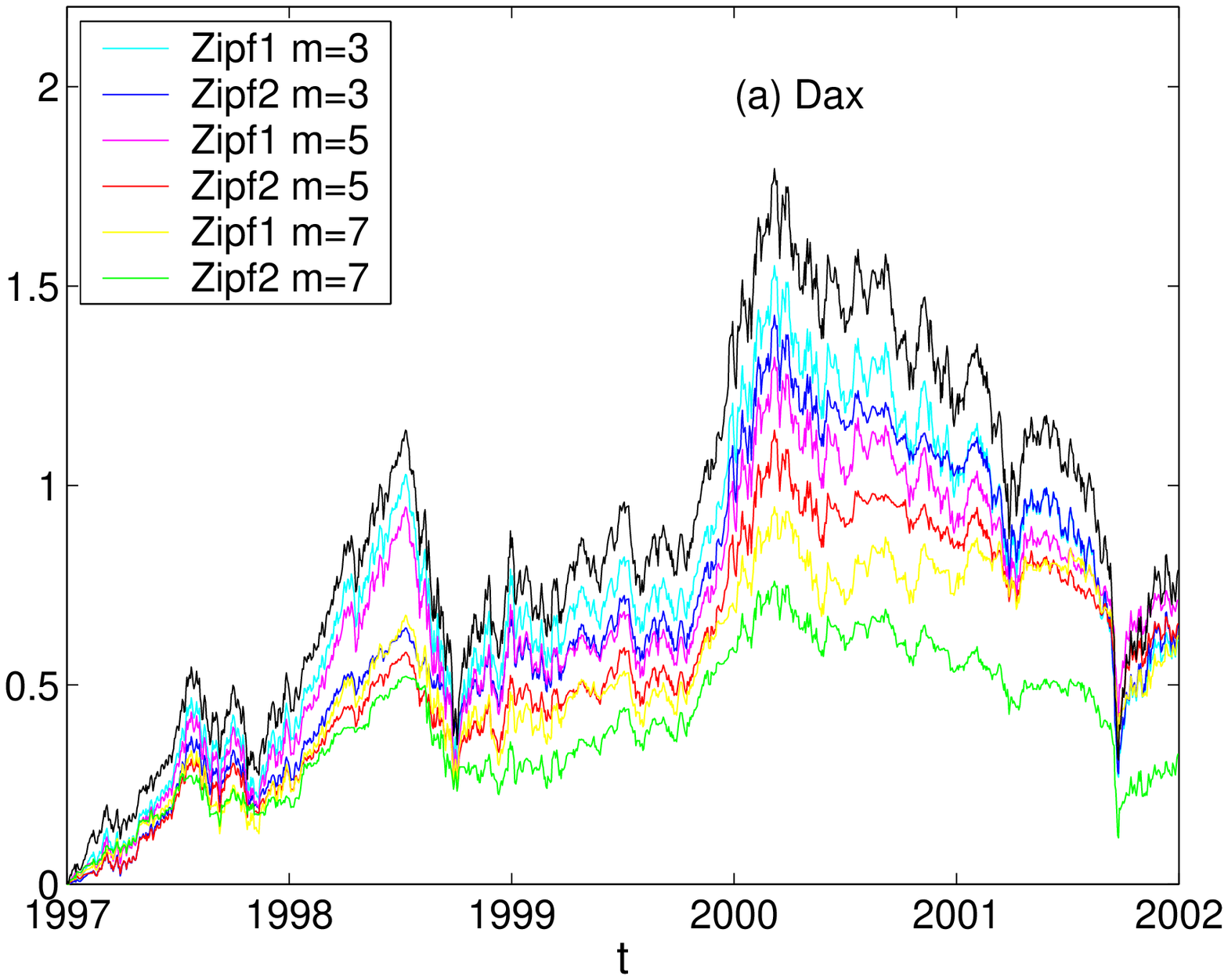}
\hfill \leavevmode \epsfysize=4cm \epsffile{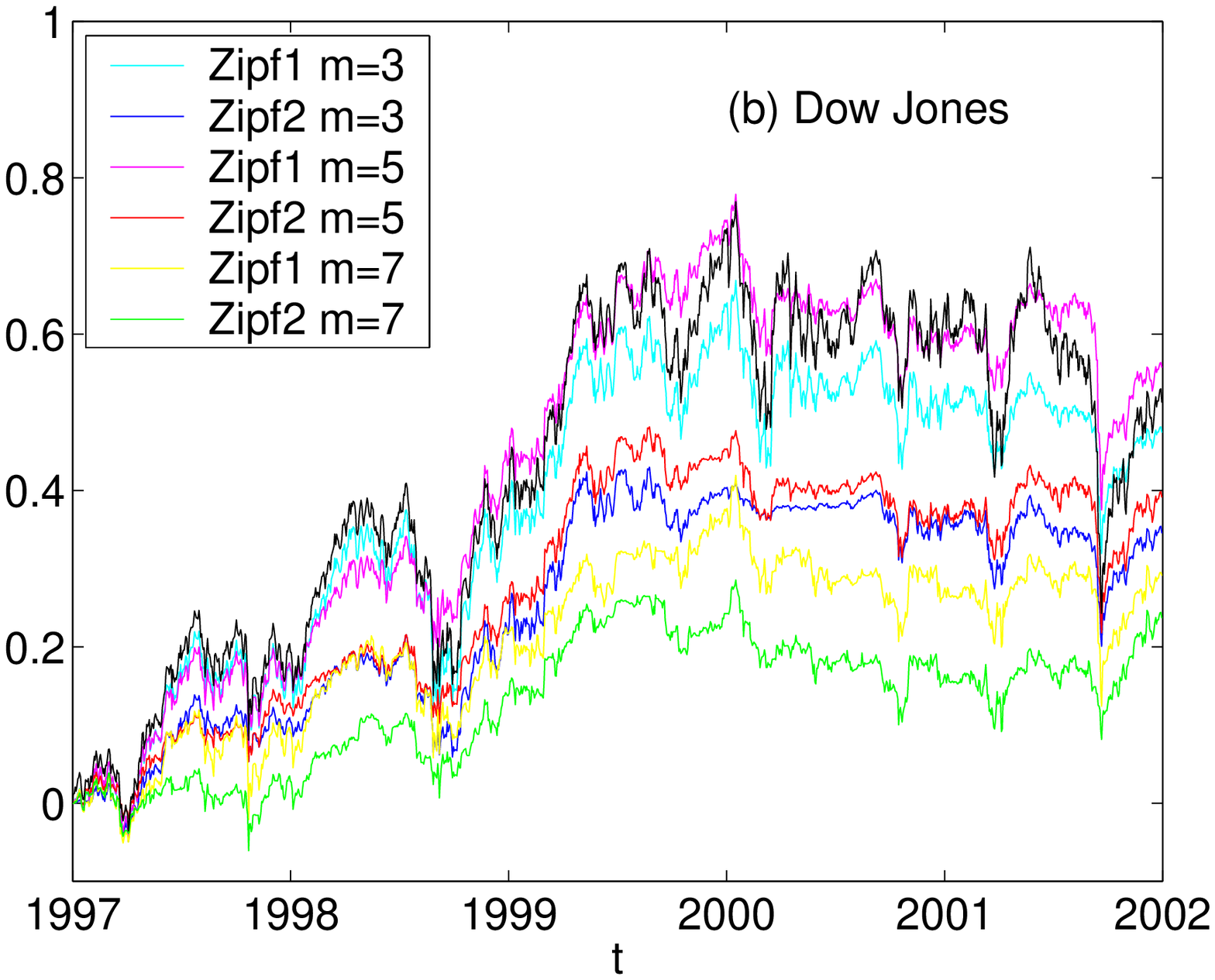} \vfill \leavevmode \epsfysize=4cm \epsffile{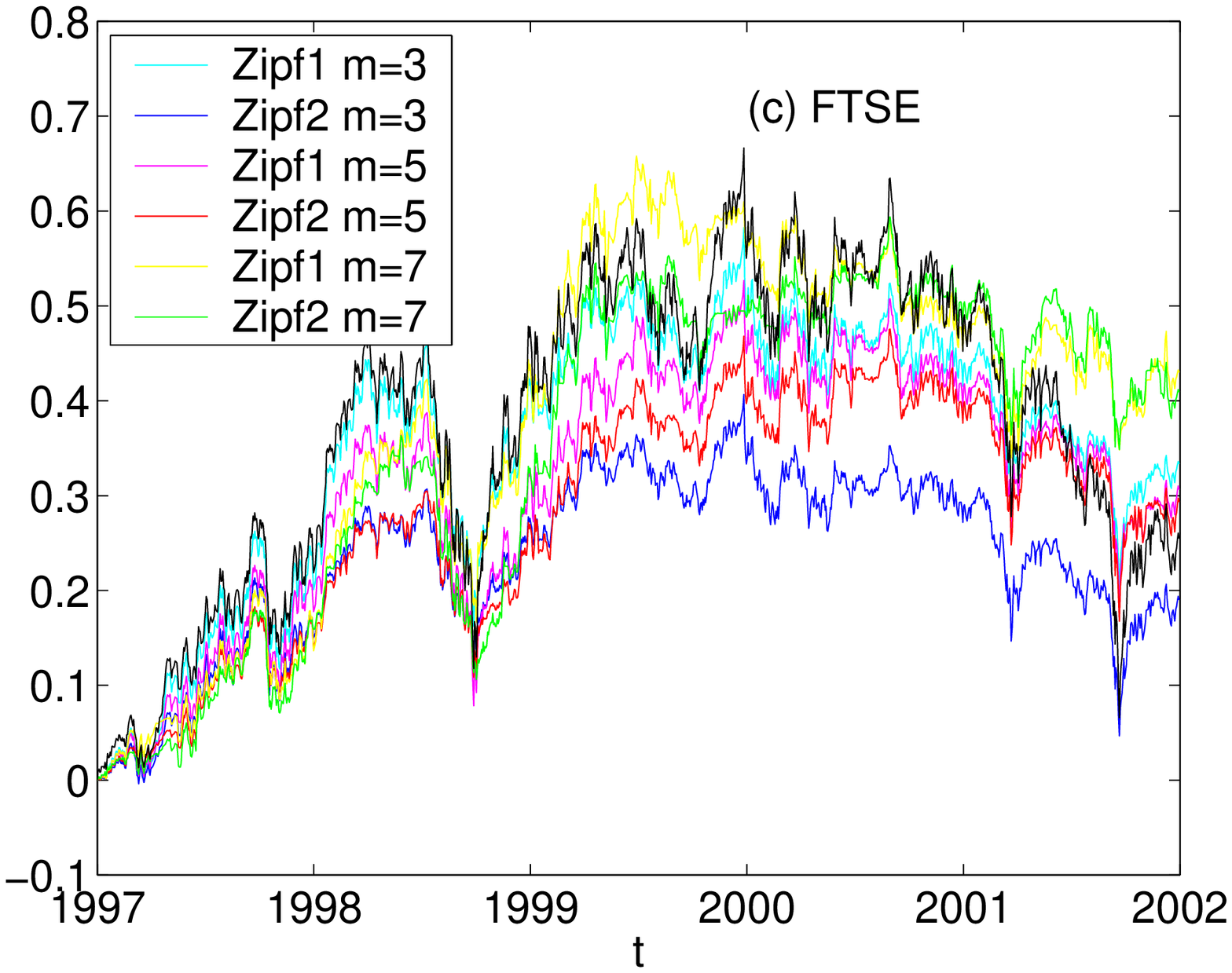} \hfill \leavevmode \epsfysize=4cm \epsffile{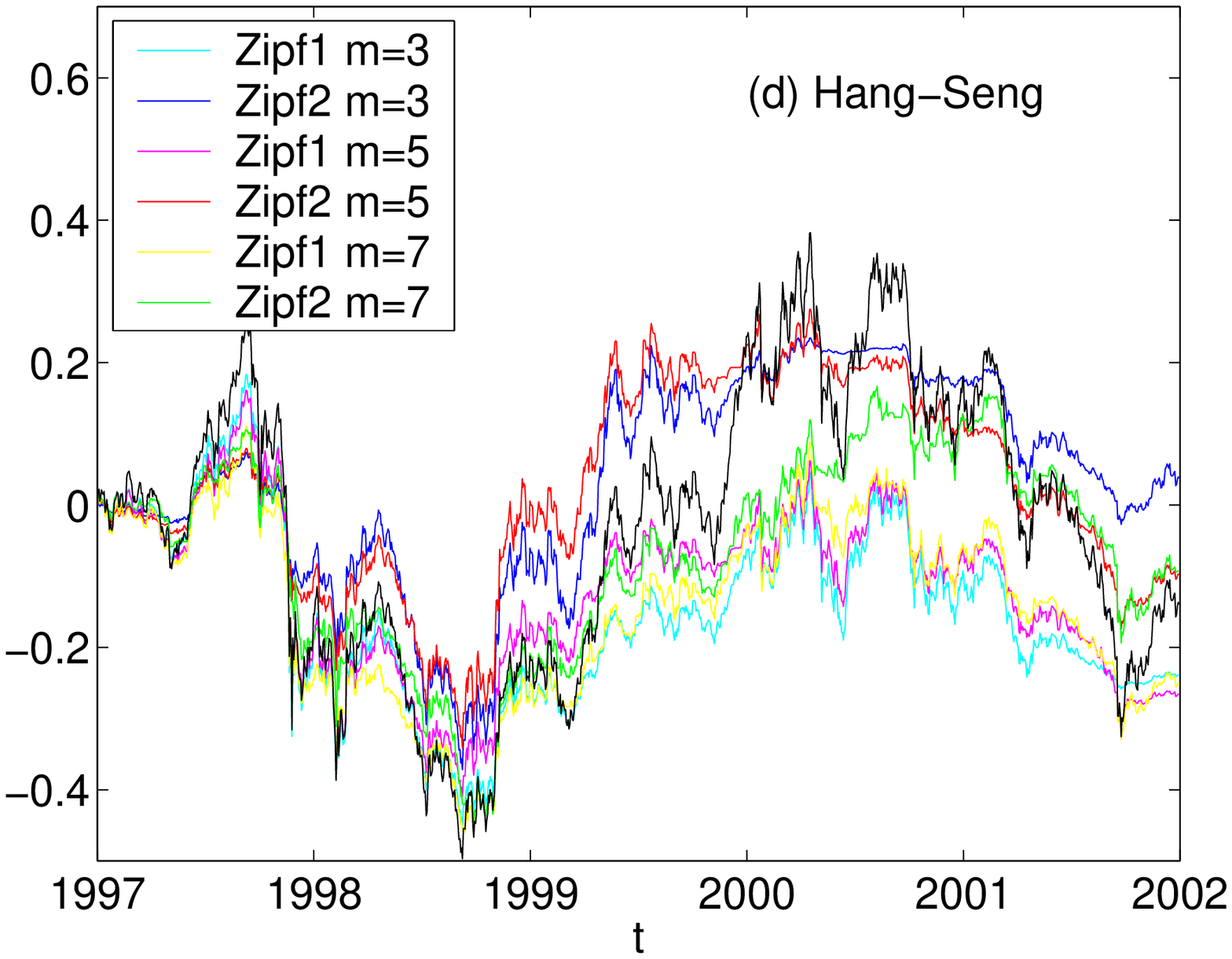} \vfill \leavevmode \epsfysize=4cm \epsffile{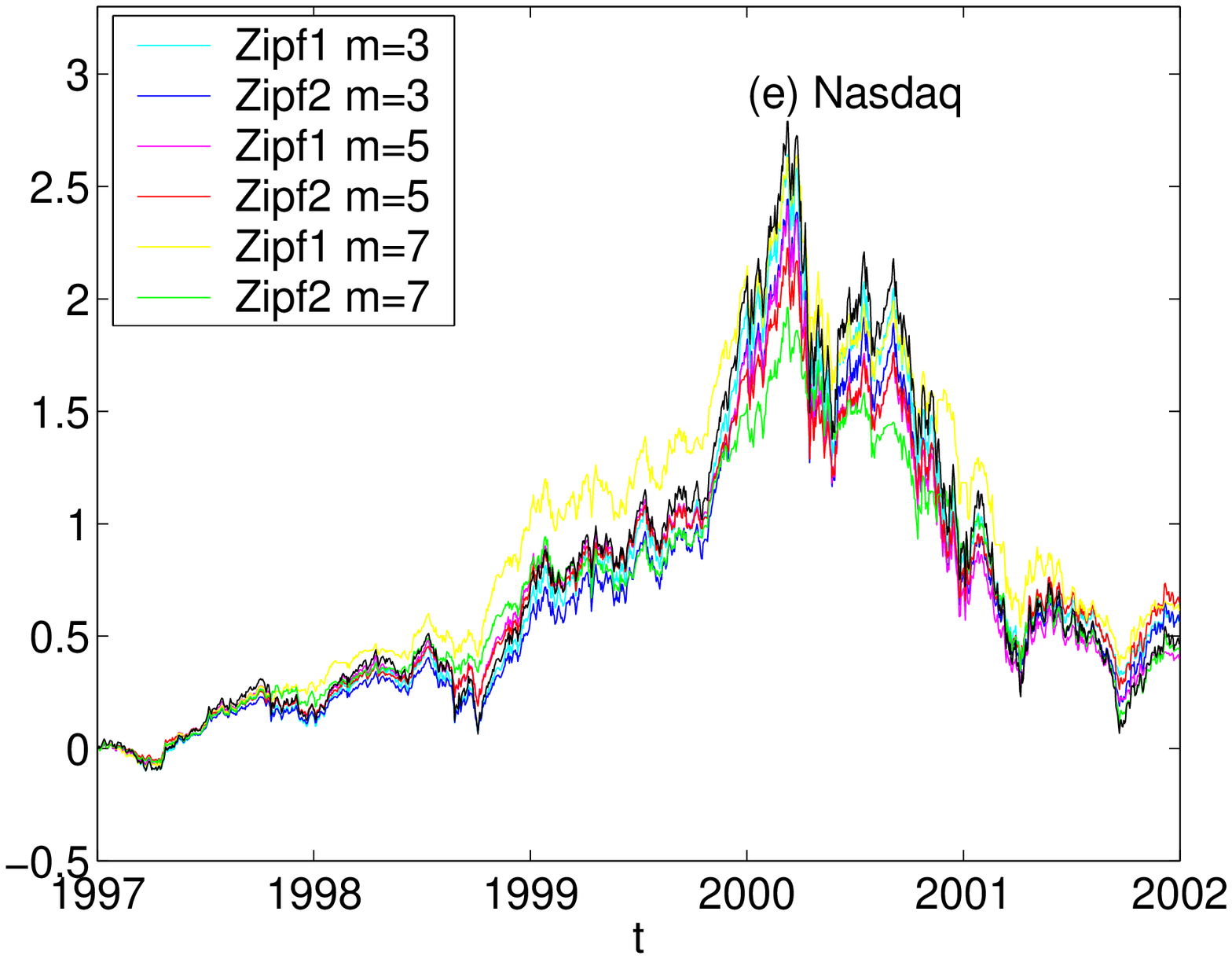}
\hfill \leavevmode \epsfysize=4cm \epsffile{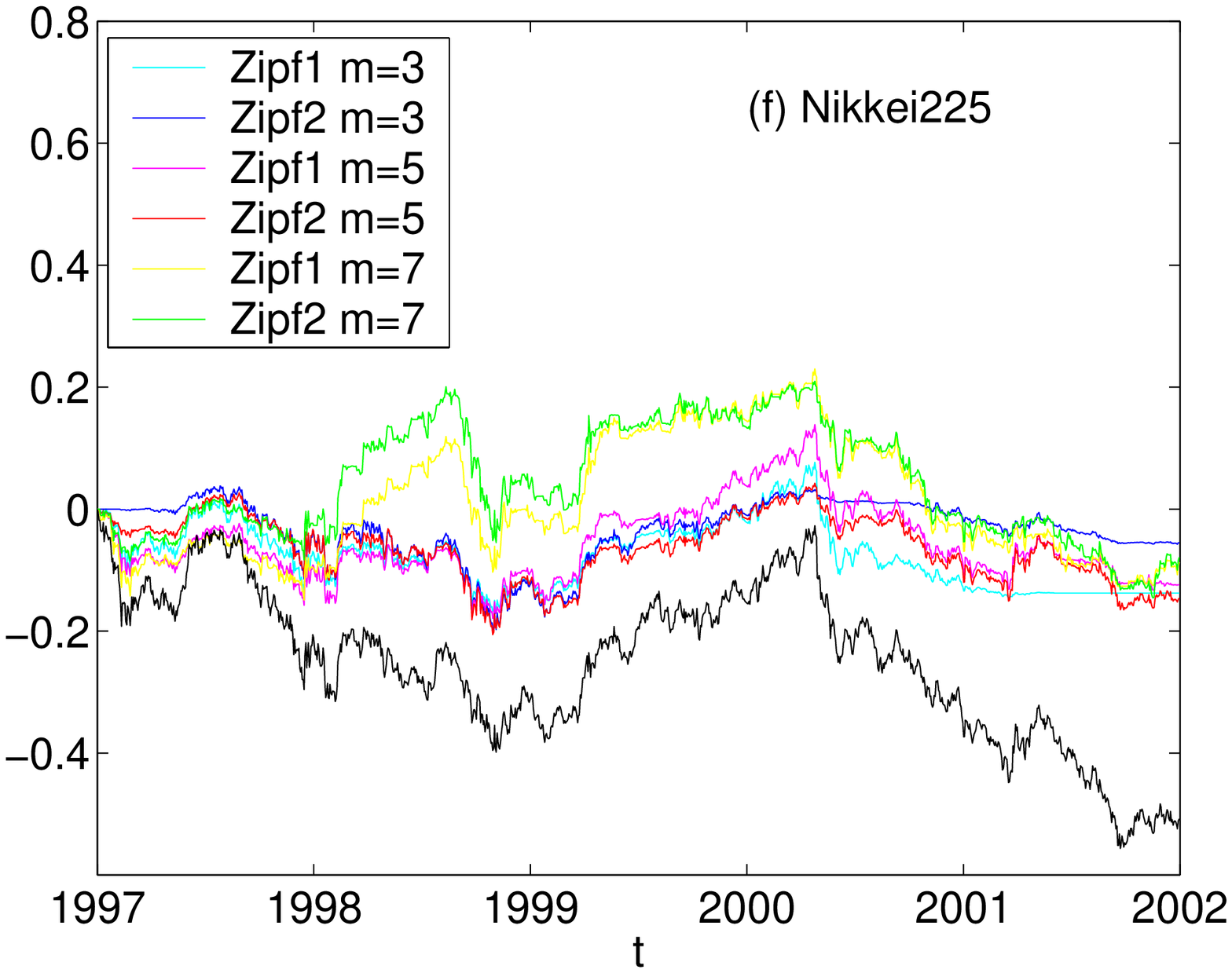} \vfill \leavevmode \epsfysize=4cm \epsffile{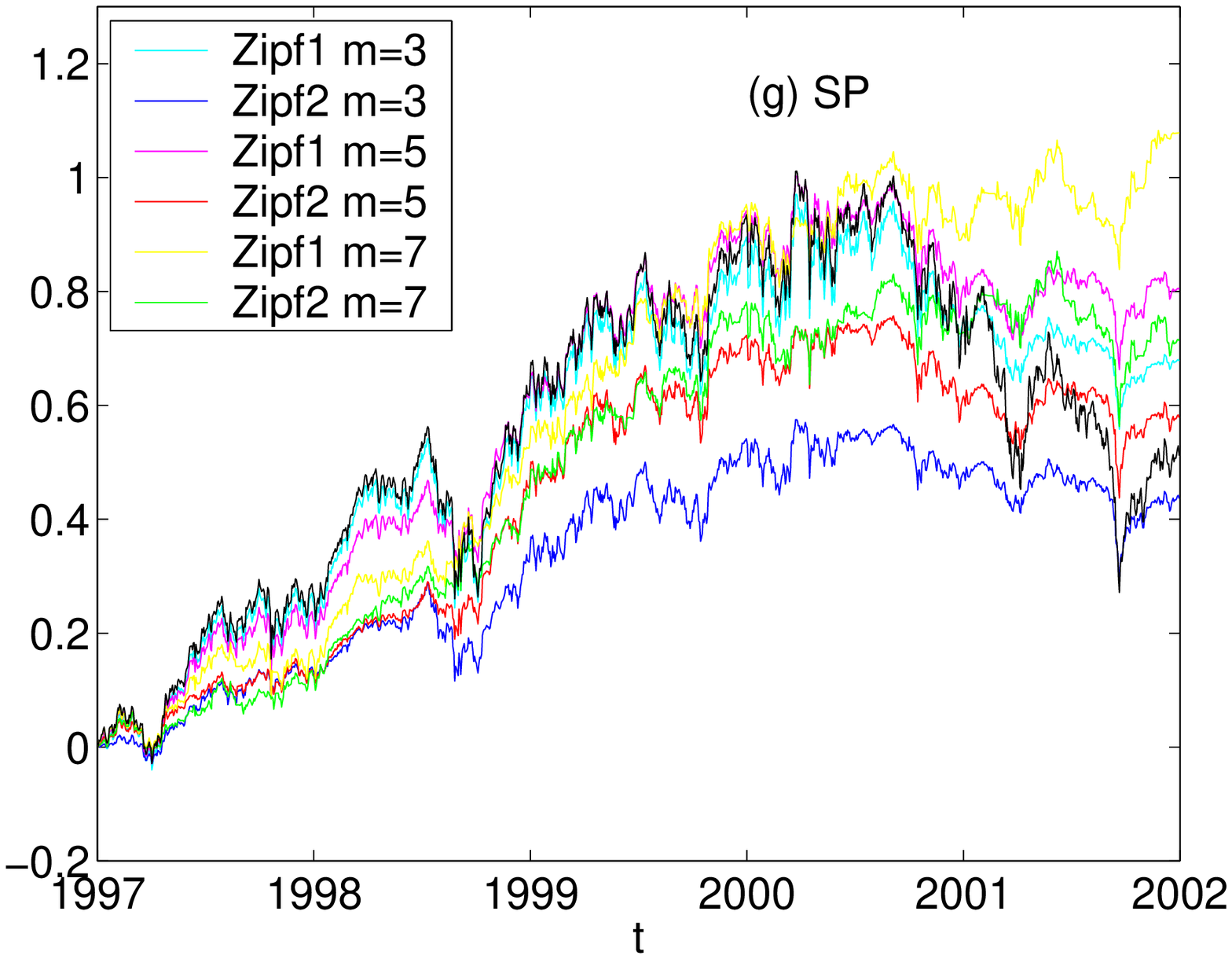} \caption{The time dependent returns for $Z1$ and $Z2$ in the case $m$=3, 5, and 7, and for $k$=2 for the seven considered
financial indices} \end{center} \end{figure} 

\section{Conclusions}

It appears (Table 2) that there is no immediate simple and general rule or universal optimum strategy. The latter depends on the volatility, i.e. the signal
roughness and the local ($m,k$)-Zipf exponent value. From the implemented simple
strategies, it occurs that "the best returns" are usually for $Z1$ with $m$ = 5,
except for NASDAQ for which a fine result arises from a	$Z1$ with $m$ = 7, (or $Z2$ and $m$=5) and for the FTSE, with either $Z1$ or $Z2$ and $m$=7. This choice
of the $m$ value and the $Z1$ strategy is conjectured to be good for large $\zeta
'$ and "non Brownian" (large $H$) cases. However for quasi-Brownian signals (and
high/low $\zeta '$ ), then it is obvious that one has reduced losses for the NIKKEI 	when one chooses a $Z1$ strategy with $m$=7; this is a very large $\zeta '$ case. This is rather similar to the FTSE case. Increased gains are found for 	$S\&P$	with $Z1$	 and $m$=7, and for DJIA 	with $Z1$
and	$m$=5, i.e. when $\zeta '$	is close to 0.1. On the contrary for the
$HS$	a $Z2$ and $m=3$ strategy should be better, i.e. for $\zeta ' <<$ 0.1. The situation is rather neutral for the DAX, the choice	$Z1$,	$m$=5 being favored.

Many other cases could be further considered, and theoretical work suggested : first one could wonder about signal stationarity. Next either a non linear (thus
like a power law) trend or a periodic background could be subtracted from the raw
signal, and the Zipf exponent time variation examined. Many other strategies are
also available.

In summary, we have translated seven financial index signals each into a text based of two letters $u$ and $d$, according to the fluctuations as in a corresponding random walk. We have calculated the Zipf exponent(s) giving the relationship between the frequency of occurrence of words of length $m<8$ made of
such ''letters'' for a binary alphabet. We have introduced considerations based on the notion of a local (or ''time'' dependent) Zipf law (and exponent). We have
imagined two simple investment strategies taking into account the linear trend of
the biased signal or not, and have reported the time dependence of the returns. 

\vskip 0.6cm

{\bf Acknowledgments}

\vskip 0.6cm

We thank K. Ivanova for stimulating discussions and comments.

{\large \bf Figure Captions} \vskip 0.5cm 

{\bf Figure 1} -- The (DAX, FTSE; DJIA, NASDAQ, S\&P500; Hang-Seng and Nikkei 225) indices have been obtained from $http://finance.yahoo.com/$. They contain $ca. $1250 data points, from Jan. 01, 1997 till Dec. 31, 2001 

\vskip 0.5cm {\bf Figure 2} -- The time dependent returns for $Z1$ and $Z2$ in the case $m$=3, 5, and 7, and for $k$=2 for the seven considered financial indices

\end{document}